\documentclass[aps,twocolumn,prl,preprintnumbers,amsmath,amssymb,superscriptaddress,floatfix,longbibliography]{revtex4}
\usepackage[pdftex]{graphicx}
\usepackage{amsmath}
\usepackage{verbatim}
\usepackage{amsmath,amssymb}
\usepackage{graphicx}   

\usepackage{hyperref}
\usepackage{indentfirst}
\usepackage{xcolor}

\begin{document}

\title{High pressure control of optical nonlinearity in the polar Weyl semimetal TaAs}

\author{Chen Li}
\affiliation{Department of Physics, California Institute of Technology, Pasadena, CA 91125, USA\looseness=-1}
\affiliation{Institute for Quantum Information and Matter, California Institute of Technology, Pasadena, CA 91125, USA\looseness=-1}
\author{Xiang Li}
\affiliation{Department of Physics, California Institute of Technology, Pasadena, CA 91125, USA\looseness=-1}
\author{T. Deshpande}
\affiliation{Department of Physics, California Institute of Technology, Pasadena, CA 91125, USA\looseness=-1}
\author{Xinwei Li}
\affiliation{Department of Physics, California Institute of Technology, Pasadena, CA 91125, USA\looseness=-1}
\affiliation{Institute for Quantum Information and Matter, California Institute of Technology, Pasadena, CA 91125, USA\looseness=-1}
\author{N. Nair}
\affiliation{Department of Physics, University of California, Berkeley, CA 94720, USA\looseness=-1}
\affiliation{Materials Sciences Division, Lawrence Berkeley National Laboratory, Berkeley, CA 94720, USA\looseness=-1}
\author{J. G. Analytis}
\affiliation{Department of Physics, University of California, Berkeley, CA 94720, USA\looseness=-1}
\affiliation{Materials Sciences Division, Lawrence Berkeley National Laboratory, Berkeley, CA 94720, USA\looseness=-1}
\author{D. M. Silevitch}
\affiliation{Department of Physics, California Institute of Technology, Pasadena, CA 91125, USA\looseness=-1}
\author{T. F. Rosenbaum}
\affiliation{Department of Physics, California Institute of Technology, Pasadena, CA 91125, USA\looseness=-1}
\author{D. Hsieh}
\email[Corresponding author]{dhsieh@caltech.edu}
\affiliation{Department of Physics, California Institute of Technology, Pasadena, CA 91125, USA\looseness=-1}
\affiliation{Institute for Quantum Information and Matter, California Institute of Technology, Pasadena, CA 91125, USA\looseness=-1}

\begin{abstract}
\noindent 
The transition metal monopnictide family of Weyl semimetals recently has been shown to exhibit anomalously strong second-order optical nonlinearity, which is theoretically attributed to a highly asymmetric polarization distribution induced by their polar structure. We experimentally test this hypothesis by measuring optical second harmonic generation (SHG) from TaAs across a pressure-tuned polar to non-polar structural phase transition. Despite the high pressure structure remaining non-centrosymmetric, the SHG yield is reduced by more than 60\% by 20 GPa as compared to the ambient pressure value. By examining the pressure dependence of distinct groups of SHG susceptibility tensor elements, we find that the yield is primarily controlled by a single element that governs the response along the polar axis. Our results confirm a connection between the polar axis and the giant optical nonlinearity of Weyl semimetals and demonstrate pressure as a means to tune this effect $in$ $situ$.   
\end{abstract}

\maketitle

The Weyl semimetal is a gapless three-dimensional phase of matter in which spin-polarized valence and conduction bands touch at isolated points \textemdash Weyl nodes \textemdash in the Brillouin zone \cite{armitage_weyl_2018,wan_topological_2011, xu_discovery_2015, lv_experimental_2015, yang_weyl_2015}. Weyl nodes serve as sinks and sources of Berry curvature in momentum space and can thus only be removed by pairwise annihilation, rendering them topologically stable. The presence of Weyl nodes, which is only allowed in the absence of time-reversal or inversion symmetry, has been shown to endow materials with exotic DC transport properties including the chiral anomaly \cite{Zyuzin_2012_chiral,zhang2016signatures,jia_weyl_2016,parameswaran_probing_2014,huang_observation_2015}, anomalous Hall  \cite{Yang_hall_2011,Burkov_2014_hall,Sun_2016_hall} and surface Fermi arc mediated cyclotron motion \cite{Moll2016, Koshino_2016_cycoltron,li2017evidence_cyclotron}. 

The transition metal monopnictide family of inversion broken Weyl semimetals has garnered additional attention for its band structure geometry induced second-order nonlinear AC response \cite{orenstein_topology_2021}. Headlined by TaAs, observations include strong light helicity-dependent injection \cite{ma_direct_2017,sirica_tracking_2019,gao_chiral_2020} and helicity-independent shift photocurrents \cite{sirica_tracking_2019,osterhoudt_colossal_2019,Ma2019}, as well as a record-setting shift current-induced optical second harmonic generation (SHG) efficiency in the near-infrared spectral range \cite{wu_giant_2017,patankar_resonance-enhanced_2018}. This SHG response has been attributed theoretically to a large skewness in the polarization distribution originating from the polar structure of these materials \cite{patankar_resonance-enhanced_2018,li2018second}. Experimentally testing this hypothesis would not only advance our fundamental understanding of the nonlinear optical properties of inversion broken Weyl semimetals, but also illuminate new pathways to control them.

\begin{figure}[b]
\centering
\includegraphics[width=3.3in]{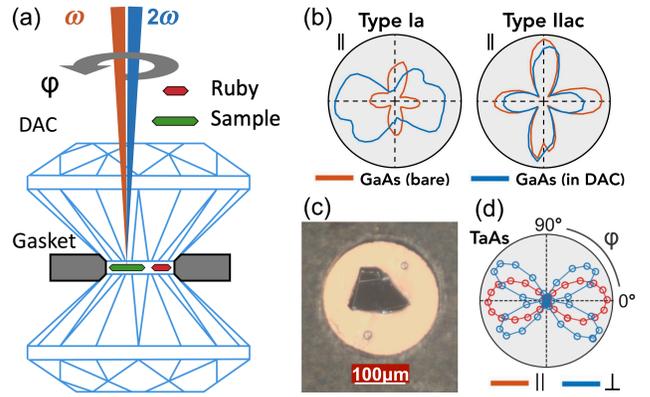}
\label{Fig1}
\caption{(a) Schematic of the diamond anvil cell (DAC) based SHG-RA setup. Incident and reflected SHG light are shown in red and blue, respectively. The angle between the polarization of the incident beam and the [1,1,-1] axis of TaAs is defined as $\varphi$. (b) Ambient pressure SHG-RA data from GaAs (001) obtained in a parallel polarization geometry through type Ia and type IIac diamonds (blue lines), compared to data taken on the bare sample outside the DAC (orange lines). \textcolor{black}{The slight asymmetries in the SHG-RA patterns are due to GaAs surface imperfections.} (c) Microscope image showing a TaAs (112) crystal sealed inside the DAC and the locations of ruby spheres for checking pressure homogeneity. (d) Ambient pressure SHG-RA data from bare TaAs (112) obtained in parallel and crossed polarization geometries.}
\end{figure}

\begin{figure*}[t]
\centering
\includegraphics[scale=0.064]{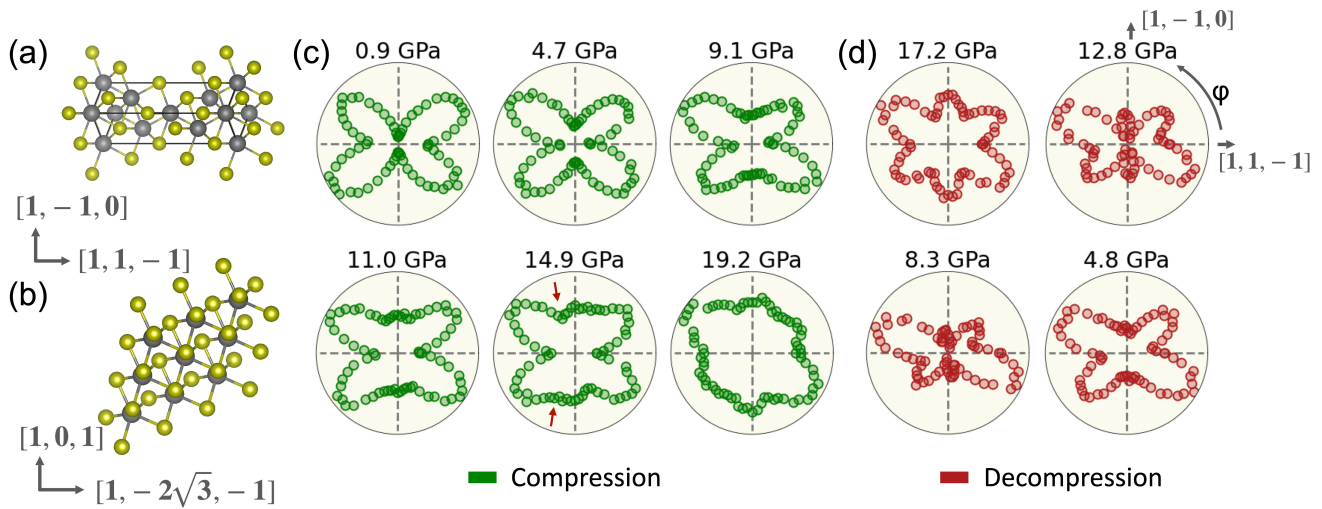}
\label{Fig2}
\caption{(a) Unit cell of ambient pressure tetragonal ($I4_{1}md$) TaAs projected onto the (112) surface. (b) Projection of high-pressure hexagonal TaAs ($P\overline{6}m2$) onto the same surface. (c) SHG-RA patterns at select pressures measured upon compression and (d) decompression after reaching a maximum pressure of 19.2 GPa. \textcolor{black}{Pressure was tuned $in$ $situ$, allowing all patterns to be acquired under identical alignment conditions.} Horizontal and vertical dashed lines lie along the [1,1,-1] and [1,-1,0] axes respectively. The loss of horizontal mirror symmetry in the 14.9 GPa data can be seen from the presence of an intensity dip at $\varphi \approx 100^{\circ}$ contrasted with hump at $\varphi \approx 260^{\circ}$ (red arrows).} 
\end{figure*}

Powder x-ray diffraction measurements on TaAs \cite{zhou_pressure-induced_2016}, supported by density functional theory calculations \cite{guo_high-pressure_2018,gupta_pressure-induced_2018,zhang_prediction_2017,lu_high-pressure_2016}, have shown that hydrostatic pressurization above a critical value $P_c$ = 14 GPa induces a phase transition from an inversion broken polar Weyl semimetal to an inversion broken non-polar Weyl semimetal. In this Letter, we present a pressure-dependent SHG rotational anisotropy (RA) study of single crystalline TaAs. We observe a structural symmetry change above $P_c$ consistent with powder x-ray diffraction data, which is accompanied by a sharp decrease in overall SHG efficiency \textemdash approaching 60\% at 20 GPa \textemdash despite the crystal symmetry remaining inversion broken. By tracking the pressure dependence of different components of the SHG susceptibility tensor, we show that this effect is dominated by the response along the polar axis, in line with a previous theoretical proposal. Moreover, we find evidence of reversibility upon slow decompression, providing a pathway to $in$ $situ$ tunable optical nonlinearity in Weyl semimetals.

Large hydrostatic pressures were applied using a 3-pin Merrill–Bassett diamond anvil cell (DAC) \cite{moggach_incorporation_2008}. Continuous $in$ $situ$ pressure tuning was implemented using a helium gas driven membrane and monitored via ruby fluorescence [Fig. 1(a)]. Use of low defect density type IIac diamond (Almax easyLab) was essential for sufficiently reducing the SHG background to allow isolation of reflected SHG light from the sample. A comparative study conducted on a GaAs test sample shows that its intrinsic SHG-RA pattern is hardly changed through a type IIac diamond, whereas it is completely obscured through a conventional type Ia diamond [Fig. 1(b)]. Single crystals of TaAs were grown by vapor transport \cite{Nair_growth_2020}. As-grown (112) surfaces were identified for SHG measurements and polished from the backside to a thickness of 20 $\mu$m. Crystals were loaded inside a pre-indented MP35N gasket pressed between two diamonds with 600 $\mu$m culets [Fig. 1(c)]. A mixture of methanol-ethanol (4:1) was used as the pressure medium. The sample was maintained at room temperature for all measurements. Light from a regeneratively amplified Ti:sapphire laser (1.5 eV photon energy, 100 kHz repetition rate) was focused at normal incidence onto a 30 $\mu$m spot (FWHM) on TaAs (112) with a fluence of 0.03 mJ/cm$^{2}$. Retro-reflected SHG light was \textcolor{black}{reflected from a dichroic beamsplitter at a 45$^{\circ}$ angle of incidence onto} a photomultiplier tube. \textcolor{black}{The difference in reflectance for $s$- versus $p$-polarized light from the dichroic beamsplitter was less than 2 \%, below our measurement noise level}. Rotational anisotropy patterns were acquired by co-rotating a pair of input and output linear polarizers \cite{TorchinskyRSI,Harter:15}. Data collected in both parallel and crossed polarization geometries under ambient pressure agree closely with previously reported results [Fig. 1(d)].

Under ambient pressure, TaAs crystallizes in a body-centered tetragonal structure. Its $I4_1md$ space group (point group 4$mm$) includes a polar axis along the $z$ direction and two mirror planes ($M_x$, $M_y$) and two glide planes ($M_{xy}$, $M_{- xy}$) that contain the $z$ axis. On the (112) surface, which is spanned by the [1,1,-1] and [1,-1,0] axes [Fig. 2(a)], the $z$ axis and $M_{xy}$ are projected onto [1,1,-1]. This results in SHG-RA patterns that are symmetric under reflection across the horizontally oriented [1,1,-1] axis [Fig. 1(d)]. The high pressure phase is characterized by a non-polar hexagonal structure (space group $P\overline{6}m2$, point group $\overline{6}m2$) that breaks $M_{xy}$ symmetry \cite{zhou_pressure-induced_2016}. Therefore, the tetragonal-to-hexagonal phase transition can in principle be detected through the loss of horizontal reflection symmetry in the SHG-RA patterns [Fig. 2(b)].   

\begin{figure}
\centering
\includegraphics[width=3.2in]{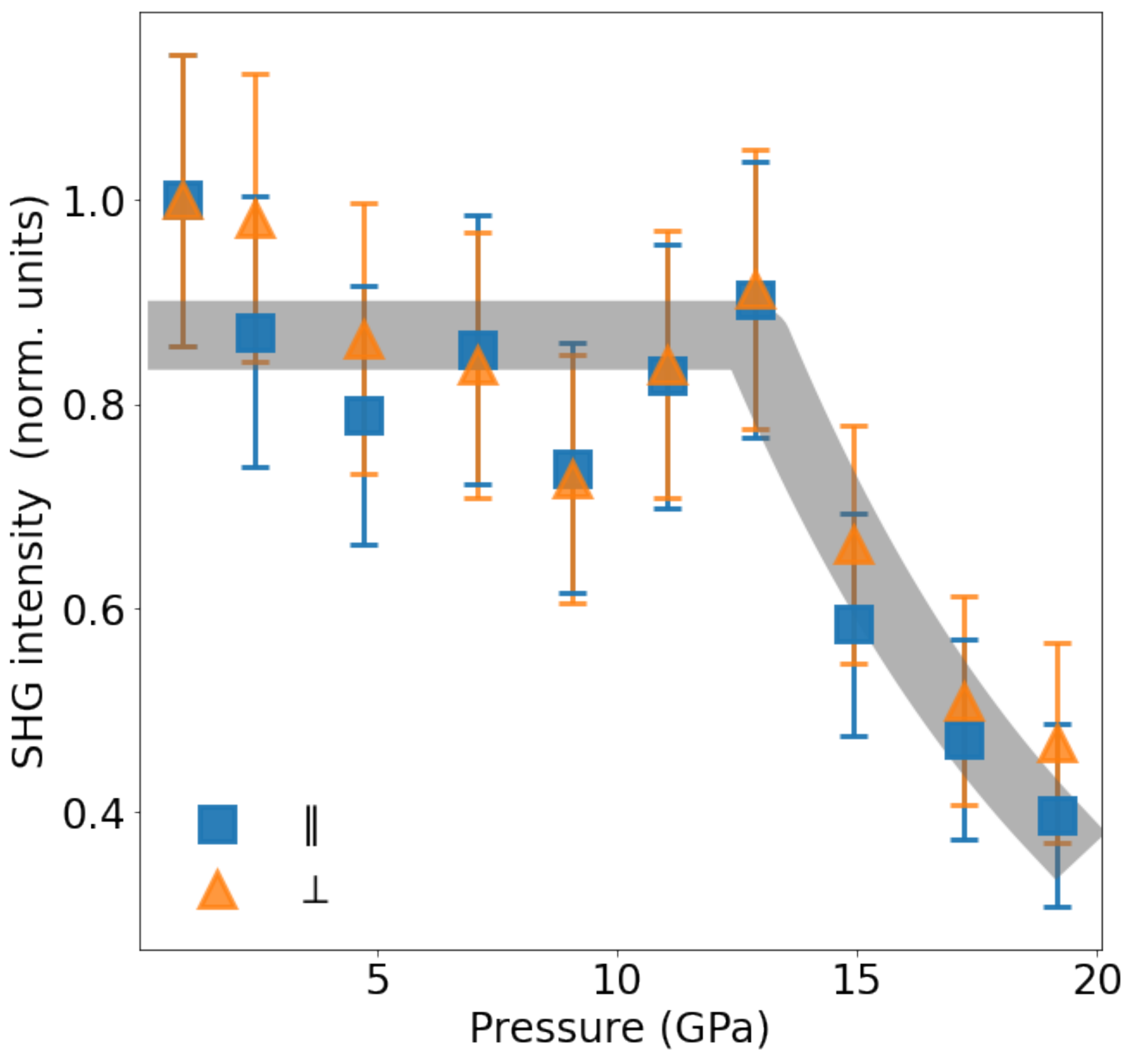}
\label{Fig3}
\caption{Pressure dependence of the $\varphi$-integrated SHG intensity from TaAs (112) measured upon compression for both parallel (blue squares) and crossed (orange triangles) polarization geometries. Each curve is normalized to its ambient pressure value. Error bars represent the intensity variation measured across different sample locations. Gray curve is a guide to the eye.}
\end{figure}

Figure 2(c) shows SHG-RA data on TaAs (112) in a crossed polarization geometry, acquired upon compression at an average rate of 1 GPa per hour. The intensities at different pressures are normalized to unity in order to better visualize the symmetry changes. At $P$ = 0.9 GPa, the sealing pressure of the DAC, the pattern is nearly indistinguishable from ambient [Fig. 1(d)]. Upon increasing pressure to 4.7 GPa, 9.1 GPa and 11 GPA, the intensity at 90$^{\circ}$ and 270$^{\circ}$ increases relative to the four major lobes, but the patterns continue to exhibit reflection symmetry about the horizontal axis. Therefore, these changes predominantly should arise from tetragonal symmetry-preserving lattice distortions. In contrast, at 14.9 GPa and 19.2 GPa the patterns become six-lobed and exhibit a clear loss of horizontal reflection symmetry, indicating a transition into the hexagonal structure. This is consistent with a previous high-pressure powder x-ray diffraction study \cite{zhou_pressure-induced_2016} that assigned $P_c$ = 14.4 GPa and found phase coexistence over a finite pressure range around $P_c$, which cannot be ruled out in our SHG data. However, whereas the x-ray diffraction study showed that the hexagonal phase persists upon decompression below $P_c$, we find that the tetragonal SHG-RA pattern is restored upon decompression, although the intensity does not recover its ambient pressure value. This demonstrates that the transition is hysteretic but possibly reversible [Fig. 2(d)]. This discrepancy may be due to a combination of our use of slow decompression over 8 hours, our lower peak pressure (19.2 GPa as opposed to 53 GPa) and differences in the behavior of single crystal versus powder samples.  

A measure of the overall SHG conversion efficiency can be obtained by integrating the SHG-RA intensity over $\varphi$ from $0$ to $2\pi$. Figure 3 shows the pressure dependence of the $\varphi$-integrated intensity in both parallel and cross polarization channels. The intensity remains essentially constant at low pressures and then starts to decrease above approximately 13 GPa, dropping to nearly 40\% of the ambient pressure value by 19.2 GPa. The fact that the intensity decrease occurs over a wide pressure window suggests a gradual transfer of population from the tetragonal to the hexagonal structure, consistent with powder x-ray diffraction measurements \cite{zhou_pressure-induced_2016}. Upon scanning the laser spot across different regions of the crystal, we observe a slight variation in intensity, which is captured by the error bars in Fig. 3. However, the overall trend versus pressure is the same. These results indicate that the large second-order optical nonlinearity of the tetragonal phase should be primarily attributed to the presence of a polar axis rather than the absence of an inversion center, which is retained in the hexagonal phase.  

To directly evaluate the contribution of the polar axis to the SHG intensity, we investigate the pressure dependence of the individual electric-dipole SHG susceptibility elements $\chi_{ijk}$, which govern the relationship between the incident electric field at frequency $\omega$ and the induced polarization at 2$\omega$ via $P_{i}^{2 \omega} = \chi_{ijk} E_{j}^{\omega} E_{k}^{\omega}$. Under normal incidence and cross-polarized geometry, the most general form of the electric-dipole SHG-RA intensity is given by $I_{\perp}(\varphi) \propto |A_{1} \cos^{3}(\varphi) + A_{2} \cos^{2}(\varphi)\sin(\varphi) + A_{3} \cos(\varphi)\sin^{2}(\varphi) + A_{4} \sin^{3}(\varphi)|^2$ where $A_1 \rightarrow A_4$ are complex numbers representing different $\chi_{ijk}$ combinations. In the case where light is normally incident on the (112) face of TaAs in the tetragonal $4mm$ phase, $A_1$ = $A_3$ = 0 by symmetry, $A_2 = \gamma (2a^{2}\chi_{zxx}+c^2\chi_{zzz}-2c^2\chi_{xzx})$ and $A_4 = \gamma (2a^{2}+c^{2})\chi_{zxx}$, where $\gamma = 1/\left(\frac{2a^{3}}{c}+ac\sqrt{2+\frac{c^{2}}{a^{2}}}\right)$ and $a$ and $c$ are the TaAs lattice constants \cite{SI}. Notably, the SHG response along the polar axis $\chi_{zzz}$ is contained in $A_2$, which was previously shown to be an order of magnitude larger than both $\chi_{zxx}$ and $\chi_{xzx}$ \cite{wu_giant_2017}. In the hexagonal $\overline{6}m2$ phase, $A_1 \rightarrow A_4$ all become symmetry allowed. 

\begin{figure*}[t]
\centering
\includegraphics[width=6.35in]{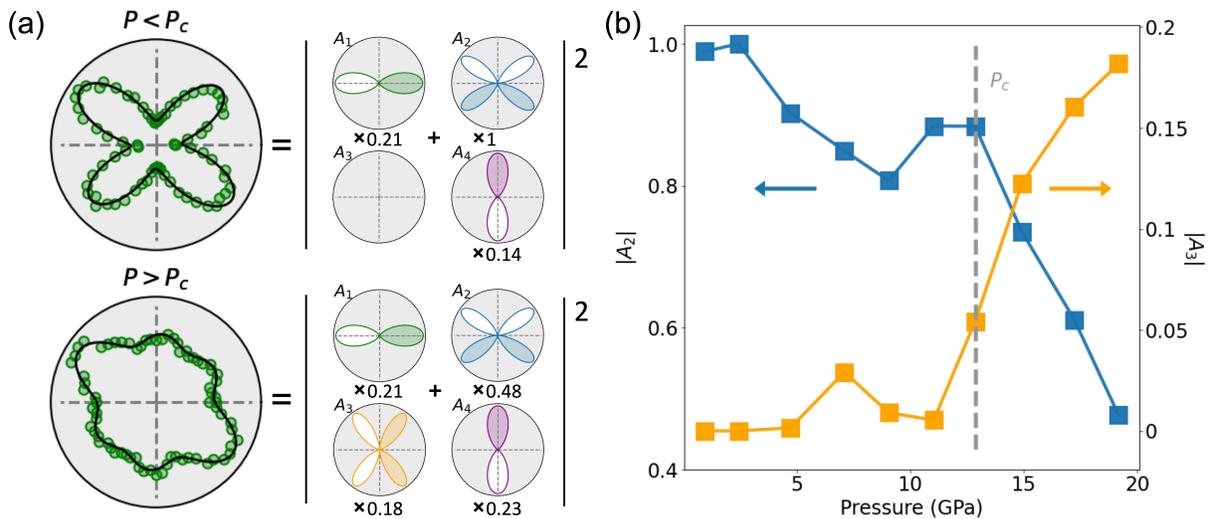}
\label{Fig4}
\caption{(a) Representative SHG-RA data acquired in crossed polarization geometry below and above $P_c$ (green circles). Fits to the expression for $I_{\perp}(\varphi)$ given in the main text are overlaid as black lines. A decomposition of the fits into its $A_1 \rightarrow A_4$ components is shown to the right. Filled versus empty lobes represent opposite signs of the associated trigonometric function. The fitted amplitude of the various terms is shown below, which are normalized to the maximum value of $A_2$ for $P < P_c$. (b) Pressure dependence of the fitted amplitudes of $A_2$ (blue) and $A_3$ (orange).}
\end{figure*}

The SHG-RA pattern at each measured pressure was fitted to the general expression for $I_{\perp}(\varphi)$. Figure 4(a) shows representative examples below and above $P_c$, which demonstrates a high quality of fit, alongside the fitted amplitudes of $A_1 \rightarrow A_4$. As expected, for $P < P_c$ the fit is dominated by $A_2$ with an order of magnitude smaller contribution from $A_4$. We also note a small contribution from $A_1$ that is present even in data collected outside the DAC \cite{SI}, possibly pointing to a slight departure of the ambient structure from $4mm$ symmetry. For $P > P_c$, all four terms become necessary in order to fit the SHG-RA patterns, with the growth of $A_1$ and $A_3$ being responsible for the loss of reflection symmetry about the horizontal axis. Moreover, the fit is no longer dominated by $A_2$, resulting in SHG-RA patterns with reduced anisotropy. The pressure dependence of $|A_2|$ and $|A_3|$ are plotted in Fig. 4(b), which can be used to approximately track the order parameters associated with the polar tetragonal phase and non-polar hexagonal phase, respectively. We find that while neither $|A_2|$ nor $|A_3|$ exhibit strong pressure dependence below $P_c$, they undergo clear downward and upward trends respectively above $P_c$, with $A_2$ falling to less than 50 \% of its ambient pressure value by 19.2 GPa. This reveals that the observed drop in overall SHG intensity above $P_c$ (Fig. 3) is predominantly due to a drastic reduction of $\chi_{zzz}$ induced by the loss of a polar axis in the hexagonal structure.  

These observations \textcolor{black}{are qualitatively consistent with} a recent theoretical proposal that relates $\chi_{zzz}$ to the third cumulant of the polarization distribution ($C_3$) \cite{patankar_resonance-enhanced_2018}. In contrast to the first moment ($C_1$), which describes the average macroscopic polarization, $C_3$ characterizes the intrinsic asymmetry of the polarization distribution, independent of the electronic center-of-mass. Although $C_3$ is generally allowed whenever inversion symmetry is broken, a simplified two-band tight-binding model of TaAs showed that it is greatly enhanced by the polar structure. Therefore, the observed decrease in $\chi_{zzz}$ above $P_c$ in our experiments can \textcolor{black}{possibly} be attributed to a drop in $C_3$ induced by the loss of a polar axis. \textcolor{black}{Verification of this hypothesis will require detailed calculations of $C_3$ for realistic models of TaAs under pressure}. By extension, our work opens a route to manipulate the large second-order nonlinearity of transition metal monopnictide Weyl semimetals by tuning their polar order parameter with hydrostatic pressure. More generally, our demonstration of high pressure SHG-RA from single crystalline TaAs provides the means to search for symmetry breaking in the vicinity of high pressure superconducting phases in non-centrosymmetric materials like TaP \cite{li2017concurrence} or Cd$_2$Re$_2$O$_7$ \cite{harter_parity-breaking_2017,malavi_mathrmcd_2mathrmre_2mathrmo_7_2016,yamaura_successive_2017}, which may be key to understanding their topological properties.

\begin{acknowledgements}
We thank Darius Torchinsky, George Rossman, Jennifer Jackson and Liang Wu for helpful discussions. High pressure optical second harmonic generation measurements, as well as construction of the instrument, were supported by the U.S. Department of Energy under Grant No. DE SC0010533. T.F.R. acknowledges support from AFOSR grant No. FA9550-20-1-0263. X.L. acknowledges support from an IQIM postdoctoral fellowship. Work by N.N. and J.G.A. was supported by the National Science Foundation under Grant No. 1905397.
\end{acknowledgements}

\newpage

\onecolumngrid

\begin{center}
    \textbf{\large Supplemental Material  \\``High pressure control of optical nonlinearity in the polar Weyl semimetal TaAs"}
\end{center}

\section{I. SHG-RA data in parallel polarization geometry}

\begin{figure*}[h]
\centering
\includegraphics[scale=0.08,clip=false]{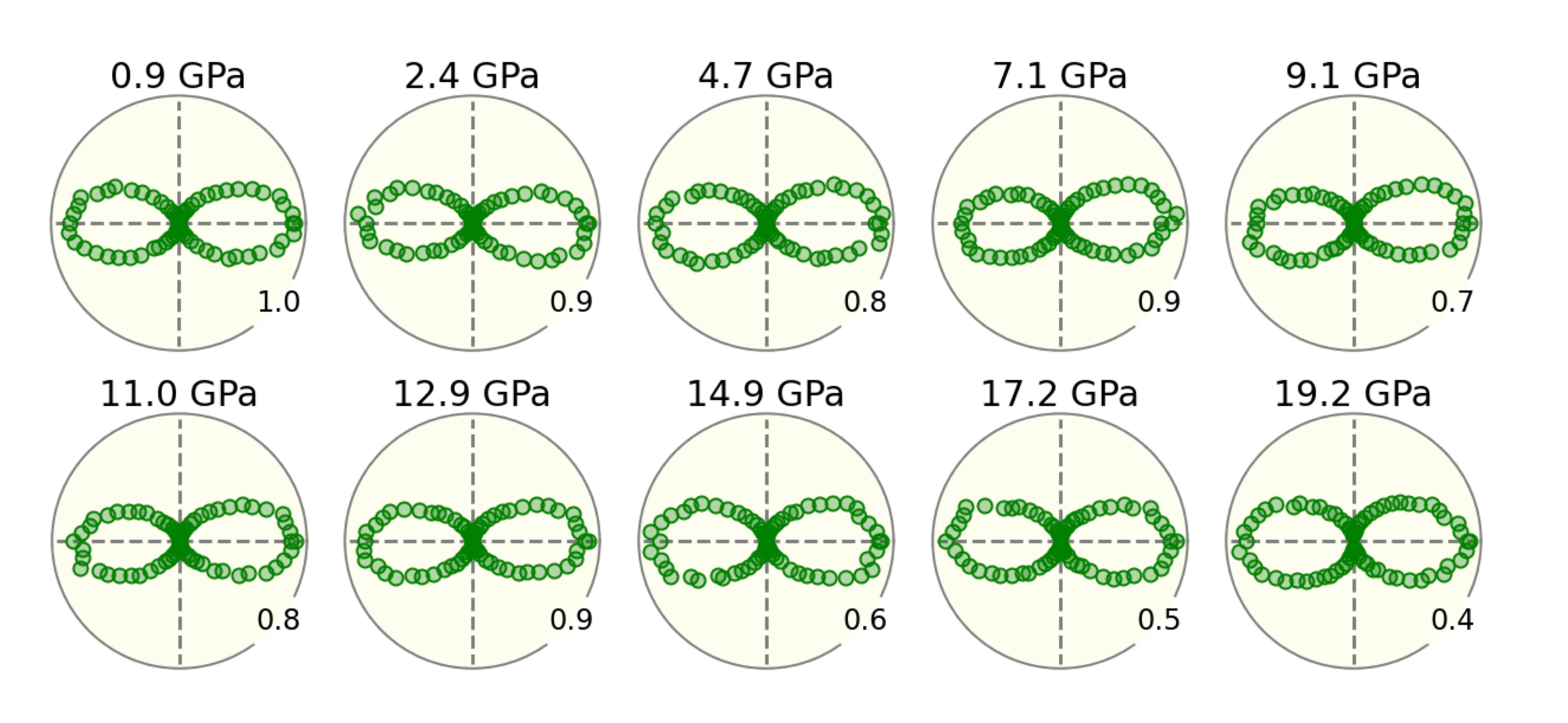}
\label{FigS2}
\caption{SHG-RA patterns at select pressure in parallel polarization geometry measured upon compression.}
\end{figure*}

The most general form of the electric-dipole SHG-RA intensity in parallel polarized geometry under normal incidence is given by $I_{\parallel}(\varphi) \propto |B_{1} \cos^{3}(\varphi) + B_{2} \cos^{2}(\varphi)\sin(\varphi) + B_{3} \cos(\varphi)\sin^{2}(\varphi) + B_{4} \sin^{3}(\varphi)|^2$, similar to the case for crossed polarized geometry described in the main text. From the (112) face of TaAs in the tetragonal $4mm$ phase, $B_2$ = $B_4$ = 0 by symmetry, $B_1 = \gamma (2a^{2}\chi_{zxx}+4a^2\chi_{xzx}+c^2\chi_{zzz})$ and $B_3 = \gamma (4a^{2}+2c^{2})\chi_{xzx}+(2a^{2}+c^{2})\chi_{zxx}$, where $\gamma = 1/\left(\frac{2a^{3}}{c}+ac\sqrt{2+\frac{c^{2}}{a^{2}}}\right)$ and $a$ and $c$ are the TaAs lattice constants. 

Figure S1 shows SHG-RA patterns at various pressures measured in parallel polarization geometry. For all pressures, the patterns are dominated by the $\chi_{zzz}$ containing $B_1$ term, which was previously shown to be an order of magnitude larger than both $\chi_{zxx}$ and $\chi_{xzx}$ \cite{wu_giant_2017_SI}. Nonetheless, careful examination of the patterns shows that the nodes at 90$^{\circ}$ and 270$^{\circ}$ are lifted at high pressure, evidencing appearance of a $B_4$ contribution that is only allowed in the hexagonal phase.

\section{II: Departure from 4$mm$ point group at ambient pressure}

\begin{figure*}[h]
\centering
\includegraphics[scale=0.07]{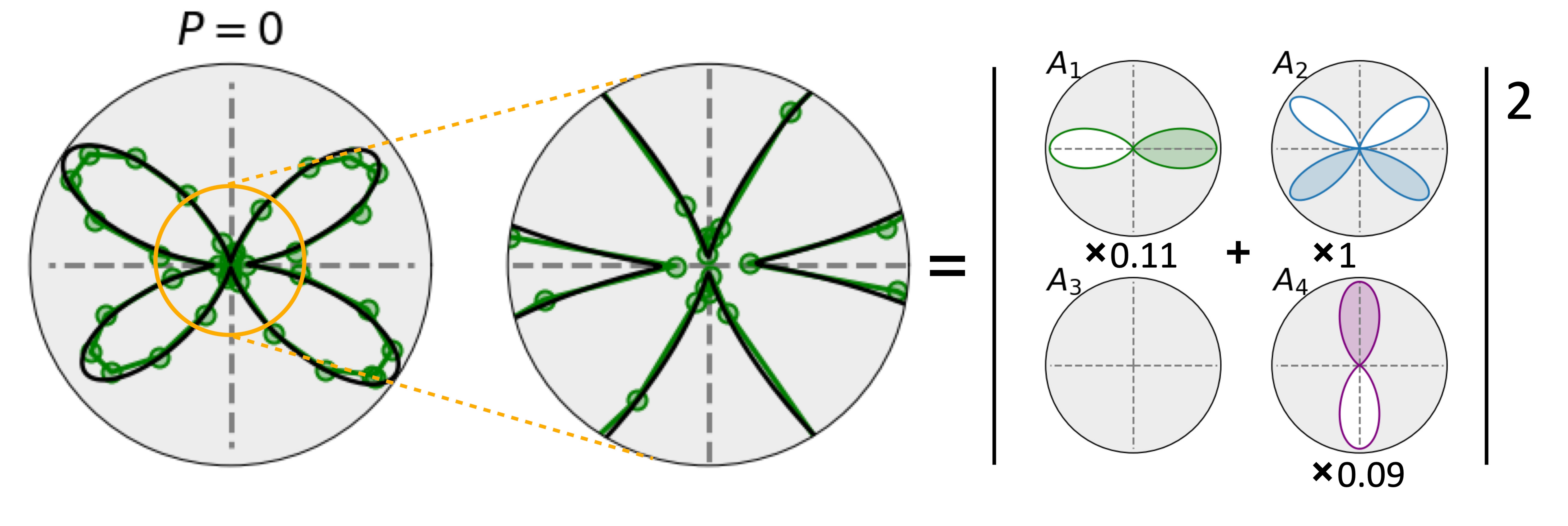}
\label{FigS3}
\caption{Typical cross-polarized SHG-RA pattern from TaAs (112) acquired at ambient pressure outside the DAC (left). A zoom-in (middle) shows a lifting of the nodes at 0$^{\circ}$ and 180$^{\circ}$, indicating a finite $A_1$ term. Fits to the general form for $I_{\perp}(\varphi)$ are overlaid as black lines. A decomposition of the fits into its $A_1 \rightarrow A_4$ components is shown to the right. Filled versus empty lobes represent opposite signs of the associated trigonometric function. The fitted amplitude of the various terms is shown below, which are normalized to $A_2$ for $P < P_c$.}
\end{figure*}

For an ideal tetragonal crystal with point group 4$mm$, the $A_1$ term in $I_{\perp}(\varphi)$ is forbidden by symmetry (see main text) and the pattern should exhibit nodes at 0$^{\circ}$, 90$^{\circ}$, 180$^{\circ}$ and 270$^{\circ}$. Yet we observe a small but finite $A_1$ component in our ambient pressure SHG-RA patterns taken outside the DAC \ref{FigS3}, manifested as a lifting of the nodes at 0$^{\circ}$ and 180$^{\circ}$. Our simulation results show that this is not due to misalignment, implying that the crystals may exhibit a slight departure from 4$mm$ symmetry.

\section{III: Pressure dependence of $A_1$ and $A_4$}

The SHG-RA patterns measured in crossed polarized geometry were fit to $I_{\perp}(\varphi) \propto |A_{1} \cos^{3}(\varphi) + A_{2} \cos^{2}(\varphi)\sin(\varphi) + A_{3} \cos(\varphi)\sin^{2}(\varphi) + A_{4} \sin^{3}(\varphi)|^2$. The fitted amplitudes of $A_2$ and $A_3$ are shown in the main text. Figure S3 shows the pressure dependence of the fitted amplitudes of $A_1$ and $A_4$. Subtle kinks appear in $|A_1|$ and $|A_4|$ around the critical pressure $P_c$. However, as expected, $|A_4|$ does not change as drastically as $|A_2|$ because $A_4$ does not contain $\chi_{zzz}$. Also, as expected, $|A_1|$ does not exhibit a clear a kink at $P_c$ as $|A_3|$ because $A_1$ is present in both the low and high pressure phases.

\begin{figure*}[h]
\centering
\includegraphics[scale=0.35]{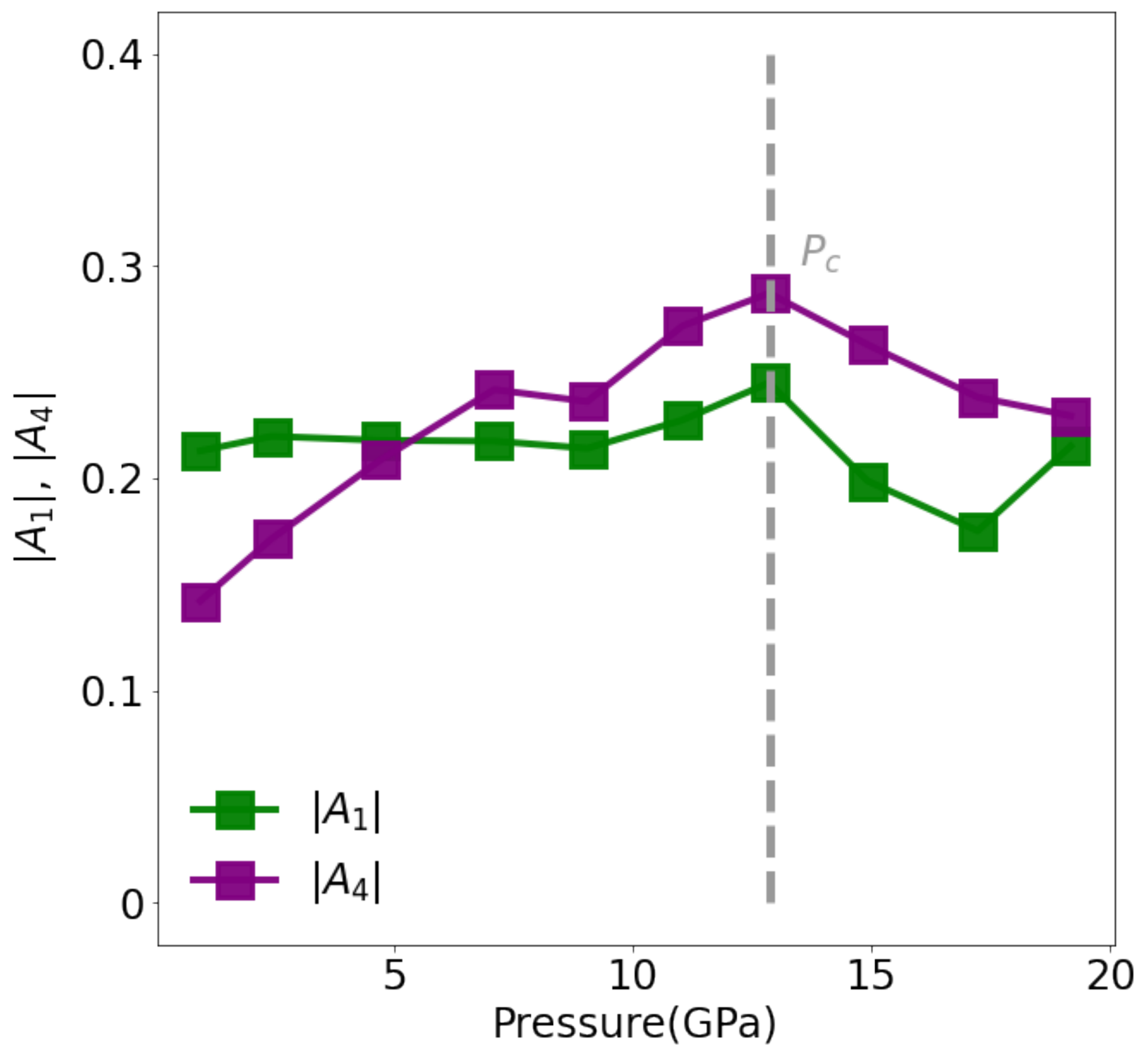}
\label{FigS4}
\caption{Pressure dependence of the fitted amplitudes of $A_1$ (green) and $A_4$ (purple) normalized to the maximum value of $A_2$ for $P < P_c$}
\end{figure*}

%

\end{document}